\documentstyle[12pt]{article}
\begin{document}
\pagestyle{plain}
\title{The polychromatic form of the Einstein photoelectric equation}
\author{Miroslav Pardy\\
Department of Physical Electronics \\
Masaryk University \\
Kotl\'{a}\v{r}sk\'{a} 2, 611 37 Brno, Czech Republic\\
e-mail:pamir@physics.muni.cz}
\date{\today}
\maketitle
\vspace{50mm}

\begin{abstract}
We generalize the Einstein photoelectric equation for the case that the matter is irradiated by the polychromatic light of blackbody,  or, by the synchrotron radiation. The work function of the   collective motion of photoelectrons is discussed.

\end{abstract}
\vspace{1cm}

\section{Introduction}

The photoelectric effect is a quantum electronic phenomenon in 
which electrons are emitted from matter after the absorption 
of energy from electromagnetic radiation.
Frequency of radiation must be above a threshold frequency, 
which is specific to the type of surface and material. 
No electrons are emitted
for radiation with a frequency below that of the threshold. 
These emitted electrons are also known as photoelectrons in this 
context. The photoelectric effect was theoretically explained by
Einstein in his paper [1] and the term "photons" was introduced for the Einstein "light quanta"  by chemist G. N. Lewis, in 1926. Einstein writes [1]: {\it In accordance with the
assumption to be considered here, the energy of light ray spreading
out from  point source is not continuously distributed over an
increasing space but consists of a finite number of energy quanta
which are localized at points in space, which move without dividing,
and which can only be produced and absorbed as complete units}.

The free electron in vacuum cannot absorb photon. It follows from the special theory of relativity. Let us prove it. If $p_{1}, p_{2}$ are the initial and final 4-momenta of electron with rest mass $m$ and $k$ is the 4-momentum of photon, then after absorption of photon by electron we write $k + p_{1} = p_{2}$, which gives when squared $k^{2} + 2kp_{1} + p^{2}_{1} = p^{2}_{2}$. Then, with $p^{2}_{1} = p^{2}_{2} = -m^{2}$ and $k^{2} = 0$, we get for the rest electron with ${\bf p}_{1} = 0$ the elementary relation $m\omega = 0$, Q.E.D..

The linear dependence on the
frequency was experimentally determined in  1915 when 
Robert Andrews Millikan showed that Einstein formula 

$$ \hbar\omega = \frac{mv^{2}}{2} + W \eqno(1)$$
was correct. Here $\hbar\omega$ is the energy of the impinging photon,
$v$ is electron velocity measured by the magnetic spectrometer and $W$ is the work function of concrete material. The work function for
Aluminium is 4.3 eV, for Beryllium 5.0 eV, for Lead 4.3 eV, for Iron
4.5 eV, and so on [2]. The  work function concerns the
surface photoelectric effect where the photon is absorbed by an
electron in a band. The theoretical determination of the work function 
is the problem of the solid state physics. On the other hand, there is the so called atomic photoeffect [3], where the ionization
energy plays the role of the work function. The system of the
ionization energies is involved in the tables of the solid
state physics. However, the work fuction of graphene [4], or, the work fuction  of the Wigner crystal was never 
determined, and it is the one of the prestige problem of the contemporary experimental and theoretical graphene physics and the Wigner crystal physics. 

In case of the volume photoeffect, the ionization work function is defined in many textbooks on quantum mechanics. Or, 

$$W = \int_{x_{1}}^{x_{2}}\left(\frac{dE}{dx}\right)dx ,\eqno(2)$$
where $E$ is the energy loss of moving electron.
 
The formula (1) is the law of conservation of energy.
The classical analogue of the equation (1) is the motion of 
the Robins ballistic pendulum in the resistive medium.

The Einstein ballistic principle is not valid inside of the
blackbody. The Brownian motion of electrons in this cavity is caused
by the repeating Compton process $\gamma + e\to \gamma + e$ and not by
the ballistic collisions. The diffusion constant for electrons must be calculated from the Compton process and not from the ballistic
process. The same is valid for electrons immersed into the cosmic relic photon sea. 
 
The idea of the existence of the Compton
effect is also involved in the Einstein article. He writes [1]: 
{\it The possibility should not be excluded, however, that electrons 
might receive their energy only in part from the light
quantum}. However, Einstein was not sure, a priori, that his idea of
such process is realistic. Only Compton proved the reality of the
Einstein statement. 

At energies $\hbar\omega < W$, the photoeffect is not realized. However, the photo-conductivity is the real process. The photoeffect is realized only  in medium and with low energy photons, but with energies $\hbar\omega > W$, which gives the Compton effect negligible. Compton effect can be realized with electrons in medium and also with electrons in vacuum. For $\hbar\omega \gg W$ the photoeffect is negligible in comparison with the Compton effect. At the same time it is necessary to say that the Feynman diagram of the Compton effect cannot be reduced to the Feynman diagram for photoeffect. In case of the high energy gamma rays, it is possible to consider the process  called photoproduction of elementary particles on protons in LHC, or photo-nuclear reactions in nuclear physics. Such processes are energetically  far from the photoelectric effect in solid state physics.  

Eq. (1) represents so called one-photon photoelectric effect, which is valid for very weak electromagnetic waves.  At
present time of the laser physics, where the strong electromagnetic
intensity is possible,  we know that so called multiphoton
photoelectric effect is possible. Then, instead of equation (1) we can
write

$$ \hbar\omega_{1} +  \hbar\omega_{2} + ...  \hbar\omega_{n} = 
\frac{mv^{2}}{2} + W .\eqno(3)$$

The time lag between the incidence
of radiation and the emission of a photoelectron is very small, 
less than $10^{-9}$ seconds. 

As na analogue of the equation (3), the multiphoton Compton effect is 
also possible: $\gamma_{1} + \gamma_{2} + ... \gamma_{n} + e  
\rightarrow \gamma + e $
and two-electron, three-electron,... n-electron  photoelectric effect is 
also possible [3]. To our knowledge the Compton process
with the entangled photons was still not discovered and elaborated. On the other hand,  there is the deep inelastic Compton effect in the high energy particle physics. 

The quantum mechanical description of the photoeffect is presented in monographs as the nonrelativistic one, or, relativistic one. In case of the thin metalic film, the light penetrates to the distance of 1000 - 10 000 atomic layers under the surface of the film. This is not valid for graphene which is the thin film with only one atomic layer [4, 5, 6]. It is posssible  to consider the photoeffect on the 2D electron system immersed into the constant magnetic field and  graphene enables to realize this project [7]. The photoelectric effect in Wigner crystals 
is the attractive problem in the solid state physics [8].

The main idea of the quantum mechanical description of the photoeffect
is that it must be described by the appropriate S-matrix element
involving the interaction of atom with the impinging photon with the
simultaneous generation of the electron. Then follows the  definition of the cross-section by the quantum mechanical equation 

$$d\sigma = \frac{2\pi}{\hbar}|V_{fi}|\delta(-I + \hbar\omega - E_{f})\frac{d^{3}p}{(2\pi)^{3}}, \eqno(4)$$ 
where $I$ is the ionization energy of an atom and $E_{f}$ is the the final energy of the emitted electron, $|V_{fi}|$ is the matrix element of the transition of electron from the initial bound state to the final state. The matrix element follows from the perturbative theory and it involves the first order term of the interaction between electron and photon. The pedagogical form of the rigorous description of the photoeffect by Stobbe [10] and Sauter [11] is presented in the theoretical textbook by Berestetzkii et al. [12], where also the equation (4) is presented.

\section{The polychromatic photoelectric equation}
     
The physical meaning of the Einstein equation (1) is in the interaction of the monochromatic photon with energy $\hbar\omega $ with an electron in matter. The possible generalization of the Einstein equation (1) is to consider the situation where the metal film absorbs the  photons with the Planckian energy distribution  of photons of the blackbody:

$$\varrho(\omega) = \frac{\omega^{2}}{\pi^{2}c^{3}}\frac{\hbar\omega}{e^{\frac{\hbar\omega}{kT}- 1}},\eqno(5)$$
or, the synchrotron radiation with the photon density [13], (in the asymptotic limit case)

$$P(\omega) = \frac{I}{\hbar\omega_{c}} \frac{9\sqrt{3}}{8\pi}\int_{y}^\infty K_{5/3}x)dx; \; y = \frac{\omega}{\omega_{c}}; \; \omega_{c} = \frac{3}{2}\left(\frac{E}{mc^{2}}\right)^{3}\frac{c}{R}\eqno(6)$$
with 

$$I = \frac{4\pi^2e^{2}\gamma^{4}}{3R}; \; \gamma = \frac{1}{\sqrt{1 - v^{2}/c^{2}}},\eqno(7)$$ 
where $R$ is the radius of the curvature, $v$ is the relativistic velocity of an electron moving along curved trajectory and $K_{5/3}$ is the modified McDonald function of the index 5/3.

In the first case with the blackbody situation, we multiply the Einstein original equation by the density of photons 

$$n(\omega) = \frac{\omega^{2}}{\pi^{2}c^{3}}\frac{1}{e^{\frac{\hbar\omega}{kT}- 1}}\eqno(8)$$
and integrate from the threshold frequency $ \omega_{0} = W/\hbar$ to infinity to get the polychromatic photoelectric equation:

$$ \int_{\omega_{0}}^{\infty}n(\omega)\hbar\omega d\omega = \int_{\omega_{0}}^{\infty}n(\omega) d\omega\frac{mv^{2}}{2} + W \int_{\omega_{0}}^{\infty}n(\omega)d\omega.\eqno(9)$$

The last equation is the generalization of the original Einstein equation from 1905 to the situation that matter is irradiated by the photons from the blackbody cavity. 

In case that the matter is irradiated by the laser field with the known spectral distribution, the the symbol $n(\omega)$ in the last equation is of the physical meaning of the spectral distribution of photons in the laser beam. 

Function 

$$ \int_{\omega_{0}}^{\infty}n(\omega) d\omega\frac{mv^{2}}{2} = E_{kin}\eqno(10)$$
has the physical meaning of the total energy of the emitted electrons of different velocities  during the photoeffect and it can be determined by the adequate experimental technique.

Function 

$$ \int_{\omega_{0}}^{\infty}n(\omega) d\omega = N(W,T)\eqno(11)$$
is the total number of photons emitted by the blackbody in the interval $(\omega_{0}, \infty)$. It depends on the work function W an on the temperature of the thermal bath which is in our case the blackbody.

We can write the polychromatic  photoelectric equation in the following form:
 
$$ \int_{0}^{\infty}n(\omega)\hbar\omega d\omega
-\int_{0}^{\omega_{0}}n(\omega)\hbar\omega d\omega = E_{kin} + W N(W, T),\eqno(12)$$
or, in the modified form 

$$ a T^{4}
-\int_{0}^{\omega_{0}}n(\omega)\hbar\omega d\omega = E_{kin} + W N(W, T),\eqno(13)$$
where the term $aT^{4}$ was obtained by the obligate mathematical procedure

$$\int_{0}^{\infty}\varrho(\omega)d\omega = \int_{0}^{\infty}n(\omega)\hbar\omega d\omega = \int_{0}^{\infty}
\frac{\omega^{2}}{\pi^{2}c^{3}}\frac{\hbar\omega}{e^{\frac{\hbar\omega}{kT}- 1}}d\omega = \frac{\pi^{2}}{15}
\frac{k^{4}T^{4}}{c^{3}\hbar^{3}} = aT^{4}.\eqno(14)$$

We know from the textbooks that 

$$a = \frac{\pi^{2}k^{4}}{15 c^{3}\hbar^{3}} = 7.56 \times 10^{-13}
{\rm erg.cm^{-3}.grad^{-4}}.\eqno(15)$$

The equation (13) is of the two scientific meaning. The first meaning is the mathematical. Namely, if we obtain from the experiment the quantity $E_{kin}$, then the equation (13) is the mathematical equation for the determination of the work function W, where, however the work function W is also inbuilt in the integral. In other words, it is the new and original  mathematical problem of elimination of some quantity from  the nontrivial equation.

The physical  meaning of the equation (13) is, that the work function W is defined by the collective motion of electrons and we know that the collective motion of electrons is not the sum of the individual motion of electrons along the individual trajectories. So, the work function obtained  from the polychromatic photoelectric equation (13) differs from the work function obtained from the monochromatic Einstein equation (1). The theoretical determination of the two different work functions represents the basic, the fundamental and the crucial problem of the quantum theory of the solid state physics and  this problem was not till this time solved. 

The same procedure can be performed using the distribution function of photons of the synchrotron radiation, where instead the 
blackbody density of photons is the synchrotron density of photons $P(\omega)$.

$$ \int_{\omega_{0}}^{\infty}P(\omega)\hbar\omega d\omega = 
E_{kin} + W N_{synchro}.\eqno(16)$$

We can easily determine the working function only by the mesurement of the total energy of the emitted electron during the photoeffect. 
 
\section{Discussion}

The article is in a some sense the extension of the Einstein original article from 1905. The main motivation of the Einstein approach was, from the historical point of view, the solid state proof of the existence of the light quanta. The possible next 
step following from the Einstein approach was the generalization of the photoelectric effect for the situation where the absorption of photons is polychromatic, generated for instance by the blackbody, or by the synchrotron. 
 
In time of the Einstein photoelectric derivation, the blackbody radiation was under discussion and the Schott formula for the synchrotron radiation was not derived. So, the Einstein motivation to go beyond his photoelectric equation was not sufficiently strong. Now, the polychromatic form of the Einstein photoelectric equation is physically meaningful.

The derived polychromatic equation can be at present time applied not only to the thin films but also to the monoatomic sheets fabricated from Carbon atoms called graphene. Application to the  bi-layer graphite, n-layer graphite and on the Wigner crystals which are spontaneously formed in graphite structures, or,  in many heterogenous structures  is not excluded.

It is also possible to consider the polychromatic photoeffect 
in the planar crystal at zero
temperature and in the very strong magnetic field. 
The motion of electrons can be approximated by the Schr\"odinger equation for an electron orbiting in magnetic field, considered for instance by author [7] . 

We do not know, a priori, how many discoveries are involved in the investigation of the polychromatic photoeffect in solid state structures, or, in  graphenes.

The work function in the Einstein equation must be determined by approriate physical measurement. In our case we use the  multiphoton photelectric equation (13), or, (16) to determine the work function. However, the result obtained from eqs. (13), or, (16) will be not identical with the result obtained form the original Einstein equation. Why? The collecive motion of electrons in multiphoton experiment generates different work function than one in case where we use only the monochromatic light generating the individual motion of electrons. The measurement  and investigation of eqs. (13), (16) can be considered as crucial and leading to the new discoveries in the photonic physics, elementary particle physics and solid state physics.

The information following from the polychromatic photoelectric effect is necessary not only in
the solid state physics, but also in the elementary particle physics in the big laboratories where multiphoton beams play the substantial role of the particle detectors.

While the last century physics growth was based on the informations following from the Einstein monochromatic photoelectric equation, the development of physics in this century will be obviously based on the polychromatic photoeffect. We hope that these perspective ideas will be considered at the universities and in the physical laboratories.

\vspace{17mm}
\noindent
{\bf References}

\vspace{7mm}
\begin{enumerate}

\item Einstein, A. Annalen der Physik {\bf 17}, (1905), 132.; The English     translation is in: AJP {\bf 33}, (1965), No. 5, May, 367. 
\item Rohlf, J. W.  Modern Physics from $\alpha$ to $Z^{0}$:
   John Willey \& Sons Inc., New York, (1994).
\item Amusia, M. Ya. Atomic photoeffect: Nauka, Moscow, (1987). (in Russian).
\item Kane, C. L. NATURE  {\bf 438}, (2005), November, 168. 
\item Dato, A.; Radmilovic, V.; Lee, Z.; Philips, J.; Frenklach, M. 
    Nano Letters {\bf 8}, No. 7, (2008),  2012.
\item Lozovik,  Yu. E.; Merkulova, S. P.; Sokolik, A. A. 
    Uspekhi Fiz Nauk {\bf 178}, No. 7, (2008),  757.
\item Pardy, M. hep-ph/0707.2668v2. 
\item Wigner, E. P. Phys. Rev. {\bf 46}, (1934), 1002. 
\item Davydov, A. S.  Quantum mechanics, Second Ed., 
     Pergamon Press, Oxford, New York, (1976).
\item Stobbe, M. Ann. der Phys. {\bf 7}, (1930), 661.
\item Sauter, F. Ann. der Phys. {\bf 9}, (1931),  217.
\item Berestetzkii, V. B.; Lifshitz, E. M.; Pitaevskii L. P.;  
     Qantum electrodynamics: Moscow, Nauka, 1989. (in Russian). 
\item Jackson, J. D. Classical Electrodynamics, 3-rd ed. John Wiley \& Sons, Inc., New York, ... (1999).
\end{enumerate}

\end{document}